\documentclass[a4paper,fleqn]{cas-dc}

\usepackage[authoryear]{natbib}
\usepackage{subcaption}
\usepackage{algorithm}
\usepackage{algpseudocode}

\def\tsc#1{\csdef{#1}{\textsc{\lowercase{#1}}\xspace}}
\tsc{WGM}
\tsc{QE}
\tsc{EP}
\tsc{PMS}
\tsc{BEC}
\tsc{DE}

\begin{document}
\let\WriteBookmarks\relax
\def\floatpagepagefraction{1}
\def\textpagefraction{.001}
\shorttitle{GHRS: Graph-based Hybrid Recommendation System}
\shortauthors{Z. Zamanzadeh Darban et~al.}

\title [mode = title]{GHRS: Graph-based Hybrid Recommendation System with Application to Movie Recommendation}

\author[1]{Zahra Zamanzadeh Darban}[orcid=0000-0003-2073-8072]
\cormark[1]
\ead{zahra.zamanzadeh@monash.edu}


\address[1]{Faculty of Information Technology, Monash University, Melbourne, Australia}

\author[2]{Mohammad Hadi Valipour}[orcid=0000-0003-1055-3955]
\ead{valipour@ostadkar.ir}


\address[2]{Department of Engineering and Product, Ostadkar Company, Tehran, Iran}

\cortext[cor1]{Corresponding author}

\begin{abstract}
Research about recommender systems emerges over the last decade and comprises valuable services to increase different companies' revenue. While most existing recommender systems rely either on a content-based approach or a collaborative approach, there are hybrid approaches that can improve recommendation accuracy using a combination of both approaches. Even though many algorithms are proposed using such methods, it is still necessary for further improvement. This paper proposes a recommender system method using a graph-based model associated with the similarity of users' ratings in combination with users' demographic and location information. By utilizing the advantages of Autoencoder feature extraction, we extract new features based on all combined attributes. Using the new set of features for clustering users, our proposed approach (GHRS) outperformed many existing recommendation algorithms on recommendation accuracy. Also, the method achieved significant results in the cold-start problem. All experiments have been performed on the MovieLens dataset due to the existence of users' side information.

\end{abstract}


\maketitle

\section{Introduction}
\label{intro}
Recommendation Systems (RS) are a type of choice advisor to overcome the explosive growth of information on the web. These systems facilitate users with personalized items (products or services), which they are more likely to be interested in. RS have been employed to a wide variety of fields: movies \citep{01wei2016a, 02m2016a}, music \citep{03mao2016a, 04horsburgh2015a}, news \citep{05shi2016a, 06wang2015a}, books, e-commerce, tourism, etc. An efficient RS may dramatically increase the number of sales of customers to boost business \citep{07jannach2010a, 08ricci2015a}. In common, recommendations are generated based on user preferences, item features, user-item interactions, and some other information such as temporal and spatial data.

RS methods are mainly categorized into Collaborative Filtering (CF), Content-Based Filtering (CBF), and hybrid recommender system based on the input data \citep{09adomavicius2005a}. CF models \citep{10salah2016a, 11polatidis2016a, 12koren2015a} aim to exploit information about the rating history of users for items to provide a personalized recommendation. In this case, if someone rated a few items, CF relies on estimating the ratings he would have given to thousands of other items by using all the other users' ratings. On the other side, CBF uses the user-item side information to estimate a new rating. For instance, user information can be age, gender, or occupation. Item information can be the movie genre(s), director(s), or the tags. CF is more applied than CBF because it only aims at the users' ratings, while CBF requires advanced processing on items to perform well \citep{13lops2011a}.

Although the CF model is preferred, it has some limitations. One of CF's limitations is known as the cold-start problem: how to recommend an item when any rating does not exist for either the user or the item? One idea to overcome this issue is to build a hybrid model by combining CF and CBF, where side information can be utilized in the training process to compensate the lack of ratings through it. Some successful approaches extend the Probabilistic Matrix Factorization \citep{14adams2010a, 15salakhutdinov2008a} to integrate side information. However, some algorithms outperform them in the general case.

There are tremendous achievements of deep learning (DL) in many applied domains in the past few decades, such as computer vision \citep{16ding2015a, 17tian2016a, 18byeon2016a, 19huang2016a} and speech tasks \citep{20graves2013a, 21xue2016a}. Deep learning models have already been studied in a wider range of applications due to its capability in solving many complex tasks. Recently, DL has been inspiring the recommendation frameworks and brought us many performance improvements to the recommender. Deep learning can capture the non-linear user-item relationships and catches the complicated relationships within the data itself from different data sources such as visual, textual, and contextual.

In recent years, the DL-based recommendation models achieve state-of-the-art recommendation tasks, and many companies apply deep learning for enhanced quality of their recommendation \citep{22covington2016a, 23sh2017a}. For example, Salakhutdinov tackled the Netflix challenge using Restricted Boltzmann Machines (RBM-CF) \citep{83salakhutdinov2007restricted, 24georgiev2013a}. AutoRec is an Autoencoder for collaborative filtering \citep{25sedhain2015a}, which uses Autoencoder to predict missing ratings. Uencoders are stacked denoising Autoencoders with sparse inputs for collaborative filtering \citep{26strub2015a}. \citet{22covington2016a} proposed a DNN-based recommendation algorithm for video recommendation on YouTube, \citet{27cheng2016a} presented an application recommender system for Google Play, and \citet{23sh2017a} presented an RNN-based recommender system for Yahoo News. All of these models have shown significant improvement over traditional models. However, the existing deep learning models have not regarded the side information about the users or items, which is highly correlative to the users' rating. Indeed, combining deep learning and side information may help us to discover a surpass solution for the considered challenges.

In this paper, we introduce a hybrid approach using Autoencoder, which tackles both challenges: learning a non-linear representation of users-items and dominating the cold start problem by integrating side information. Compared to previous models in that direction \citep{25sedhain2015a, 26strub2015a, 28wu2016a}, our framework integrates the users' preferences, similarities, and side information in a unique matrix. This conjunction leads to improved results in CF.

The outline of the paper is organized as follows. First, Section \ref{relatedworks} discusses related works in both Autoencoder-based and hybrid recommendation models. Then, our proposed model is described in Section \ref{ghrs}. Finally, experimental results are given and discussed in Section \ref{testandres} and followed by a conclusion section.

\section{Related Works}
\label{relatedworks}
This section introduces the categories of DL-based recommendation models and then focuses on advanced research to identify the most outstanding and promising progress in recent years.

\subsection{Deep Learning based Recommendation Models}
Deep learning is a research field of machine learning. It learns multiple levels of representations and abstractions from data and it can solve both supervised and unsupervised learning tasks. We can categorize the existing recommendation models based on the types of employed deep learning approaches into the following two classes \citep{29zhang2019a}

\begin{itemize}

	\item The recommendation with Neural Building Blocks; In this category, the deep learning technique determines the recommendation model's applicability. For example, MLP can simply model the non-linear interactions between users and items; CNNs can extract local and global representations from heterogeneous data sources like text and image; recommender system can model the temporal dynamics and sequential evolution of content information using RNNs.
	\item The recommendation with Deep Hybrid Models; Some DL-based recommendation models utilize more than one deep learning technique. Deep neural networks' flexibility makes it possible to combine several neural building blocks to complement one another and form a more powerful hybrid model. There are many possible combinations of these deep learning techniques, but not all have been exploited.
\end{itemize}

Additionally, we review and summarize some publications which utilize Autoencoder, and they will be discussed in the following sub-sections.

\subsection{Autoencoder based Recommendation Models}
\begin{figure}
	\centering
	\includegraphics[scale=0.35]{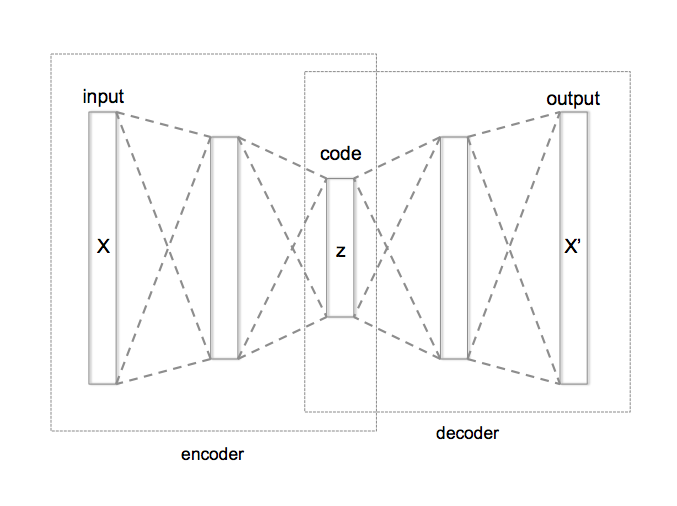}
	\caption{Schematic structure of an Autoencoder with Three fully connected hidden layers. The code (z) is the most internal layer.}
	\label{FIG:01}
\end{figure}

Autoencoder is an unsupervised model attempting to reconstruct its input data in the output layer. In general, the bottleneck layer (the middle-most layer) is used as a salient feature representation of the input data \citep{29zhang2019a}. The schematic of basic Autoencoder is illustrated in Figure \ref{FIG:01}, which output $X'$ should become closer to the input $X$ and the bottleneck layer is shown by $z$. The main variants of Autoencoders can be considered as denoising Autoencoder, marginalized denoising Autoencoder, sparse Autoencoder, contractive Autoencoder and variational Autoencoder \citep{50goodfellow2016a}.

There are two general ways to apply Autoencoder to a recommender system \citep{29zhang2019a}:
\begin{enumerate}
	\item Using Autoencoder to learn lower-dimensional feature representations at the bottleneck layer; or
	\item Filling the blanks of the interaction matrix directly in the reconstruction layer.

\end{enumerate}

Almost all the Autoencoder variants such as denoising Autoencoder, variational Autoencoder, contractive Autoencoder, and marginalized Autoencoder can be applied to the recommendation task. In this paper, we employed the first technique to extract new low-dimension features. Figure 1 illustrates the structure of different recommendation models based on Autoencoder \citep{29zhang2019a}.

\subsubsection{Autoencoder based Collaborative Filtering Models}
One of the successful applications is to consider collaborative filtering from the Autoencoder perspective. \textit{AutoRec} \citep{25sedhain2015a} took user partial vectors $r^{(u)}$ or item partial vectors $r^{(i)}$ as input and attempted to reconstruct them in the output layer. Indeed, it has two variants: item-based AutoRec (I-AutoRec) and user-based AutoRec (U-AutoRec), corresponding to the two types of inputs.

There are essential points about AutoRec that worth noticing \citep{29zhang2019a}. First, I-AutoRec performs better than U-AutoRec, which may be due to the higher variance of user partially observed vectors. Second, a different combination of activation functions will influence the performance significantly. Third, moderately increasing the hidden unit size will improve the result as expanding the hidden layer dimensionality gives AutoRec more capacity to model the input characteristics. Furthermore, adding more layers to formulate a deep network can lead to slight improvement.

\begin{figure}

	\begin{subfigure}{.5\textwidth}
		\centering
		\includegraphics[width=0.6\linewidth]{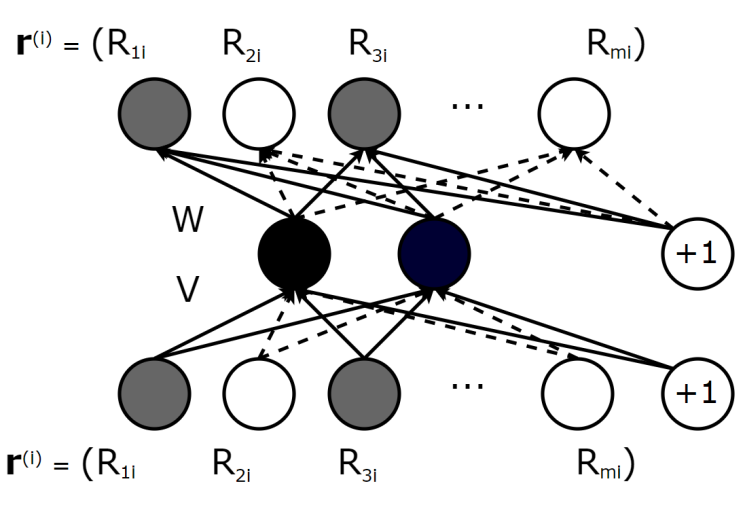}
		\caption{Item based AutoRec}
		\label{FIG:02:a}
	\end{subfigure}

	\begin{subfigure}{.5\textwidth}
		\centering
		\includegraphics[width=0.7\linewidth]{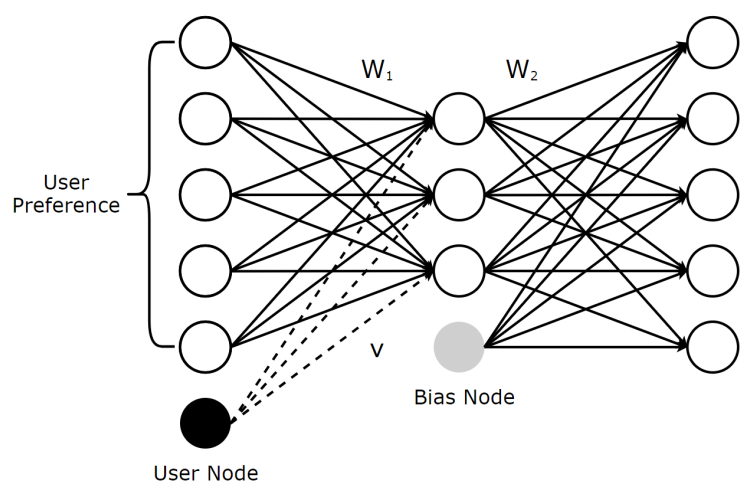}
		\caption{Collaborative denoising Autoencoder}
		\label{FIG:02:b}
	\end{subfigure}

	\begin{subfigure}{.5\textwidth}
		\centering
		\includegraphics[width=0.7\linewidth]{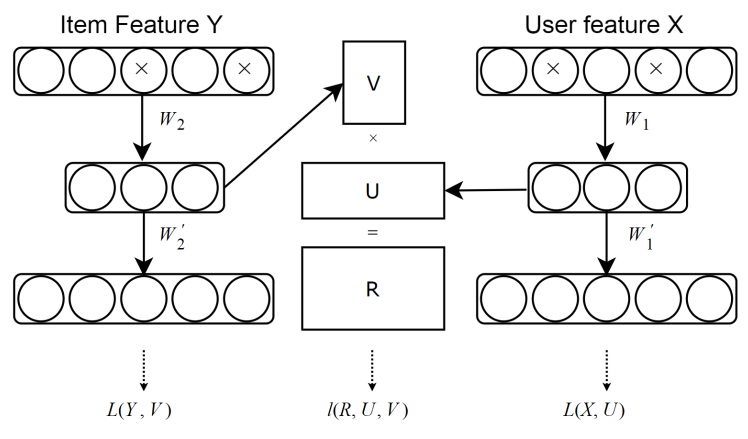}
		\caption{Deep collaborative filtering}
		\label{FIG:02:c}
	\end{subfigure}

	\caption{Illustration of: (a) Item based AutoRec; (b) Collaborative denoising Autoencoder; (c) Deep collaborative filtering framework \citep{29zhang2019a}}
	\label{FIG:02}
\end{figure}

CFN \citep{30strub2016a, 26strub2015a} is a continuation of AutoRec, and posses the following two improvements:

\begin{enumerate}
	\item Deploying the denoising techniques makes CFN more robust.
	\item Incorporating the side information such as user profiles and item descriptions mitigates the sparsity and cold start influence.
\end{enumerate}

The CFN input is also partially observed vectors, so it also has two variants: I-CFN and U-CFN, taking $r^{(i)}$ and $r^{(u)}$ as input, respectively. Masking noise is imposed as a great regularizer to better deal with missing elements (with zero value). Further extension of CFN also incorporates side information. However, instead of just integrating side information in the first layer, CFN injects side information in every layer \citep{29zhang2019a}.

Collaborative Denoising Autoencoder (CDAE). The three models reviewed earlier are mainly designed for rating prediction, while CDAE \citep{28wu2016a} is principally used for ranking prediction. The input of CDAE is user partially observed implicit feedbacks. If the user likes the movie, the entry value is one, otherwise zero. It can also be considered as a preference vector that reflects the user's interests in items \citep{29zhang2019a}. Figure 1b illustrates the structure of CDAE.

This model uses a unique weight matrix for each user and has a notable impact on model performance. Parameters of CDAE are also learned by minimizing the reconstruction error. CDAE initially updates its parameters using SGD overall feedbacks. However, it is impractical to consider all ratings in real-world applications. A negative sampling technique has been proposed to sample a small subset from the negative set (items with which the user has not interacted), which reduces the time complexity substantially without degrading the ranking quality \citep{29zhang2019a}.

Muli-VAE and Multi-DAE \citep{32liang2018a} proposed a variant of variational Autoencoder for recommendation with implicit data, showing better performance than CDAE. These methods introduced a principled Bayesian inference approach for parameter estimation and showed agreeable results than generally used likelihood functions.

Based on a survey by \citet{29zhang2019a}, Autoencoder-based Collaborative Filtering (ACF) \citep{33ouyang2014a} is the first Autoencoder based collaborative recommendation model. Instead of using the original partial observed vectors, it decomposes them by integer ratings. Like AutoRec and CFN, ACF aims at reducing the mean squared error as the cost function. But, there are two demerits of ACF; it loses to deal with non-integer ratings, and the decomposition of partially observed vectors increases the sparseness of input data and drives to worse prediction accuracy.

\subsubsection{Feature Representation Learning with Autoencoder}
Autoencoder is a dominant feature representation learning approach, and it can be used in recommender systems to learn feature representations from users-items content features. In the following, we will summarize some of the related methods.

Collaborative Deep Learning (CDL) \citep{34wang2015b} is a hierarchical Bayesian model that integrates stacked denoising Autoencoder (SDAE) into probabilistic matrix factorization. The method proposed a general Bayesian deep learning framework \citep{35wang2016a} to combine the deep learning and recommendation model. The framework consists of two tightly coupled parts: the perception component (deep neural network) and task-specific component. Mainly, CDL's perception component is a probabilistic representation of ordinal SDAE, and PMF (Probability Mass Function) works as the task-specific component. This tight combination enables CDL to balance the impacts of side information and interaction records.

Collaborative Deep Ranking (CDR). CDR \citep{38ying2016a} is devised specifically in a pairwise framework for top-n recommendation. Some studies have demonstrated that the pairwise model is more suitable for ranking lists generation. Experimental results also show that CDR outperforms CDL in terms of ranking prediction \citep{29zhang2019a}.

Deep Collaborative Filtering Framework is a general framework for unifying deep learning approaches with a collaborative filtering model \citep{39li2015a}. This framework makes it easy to utilize deep feature learning techniques to build hybrid collaborative models \citep{29zhang2019a}. There is a marginalized denoising Autoencoder-based collaborative filtering model (mDA-CF) on top of this framework. In comparison to CDL, mDA-CF explores more computationally efficient variants of the Autoencoder. The method saves the computational costs of searching sufficient corrupted input by marginalizing the corrupted input, and it makes the mDA-CF more scalable than CDL. Plus, mDA-CF embeds content information of items and users, while CDL only regards item features' effects.

AutoSVD++ \citep{40zhang2017a} uses contractive Autoencoder \citep{41rifai2011a} to learn item feature representations, then integrates them into the classic recommendation model, SVD++. The model posses the following advantages \citep{29zhang2019a}
\begin{enumerate}
	\item Compared to other Autoencoder variants, contractive Autoencoder captures the infinitesimal input variations.
	\item It models the implicit feedback to enhance the accuracy further.
	\item An efficient training algorithm is designed to reduce training time.
\end{enumerate}

HRCD \citep{42wei2017a} is a hybrid collaborative model based on Autoencoder and timeSVD++. It is a time-aware model that uses SDAE to learn item representations from raw features and solve the cold item problem \citep{29zhang2019a}.

\section{Graph-based Hybrid Recommendation System}
\label{ghrs}
In this section, we focus on our proposed method which can be categorized as a hybrid recommendation system. First we define the basic notations used throughout the paper. Next, we describe the proposed model in an architectural view and algorithmic steps. Then, graph-based features will be declared separately. Finally, we will explain about the clustering method and how we find the optimum number of clusters.

We first define the basic notations used throughout this paper. Given the set of n users, $U = \{u_1, ... , u_n\}$ , and the set of m item, $I = \{i_1, ... , i_m\}$, all user-item pairs can be denoted by an n-by-m matrix $R = U \times I$ , where the entry $r_{ui}$  indicates the assigned value of implicit feedback of user $u$ to item $i$. If $r_{ui}$ has been observed (or known), it is represented by a specified rating associated in a specific range and interval; otherwise, a global default rating is zero. We used this matrix to find similarity between users' preferences. After generating the similarity graph which represents users as nodes and the relations as edges, we extract the features from this graph, $F_g = \{f_1, ... , f_g\}$, and preserve them in the n-by-g matrix. We collect some users' features from the dataset, which are called side information, $F_u = \{f_1, ... , f_u\}$, some items' side information $F_i = \{f_1, ... , f_i\}$ and obtain the combined feature matrix which is n-by-g+s. Without loss of generality, we categorized all the features (both graph features and side information) as binary which enlarged final feature vector for each user.

\subsection{Architecture}
\label{architecture}
\begin{figure*}
	\centering
	\includegraphics[scale=0.29]{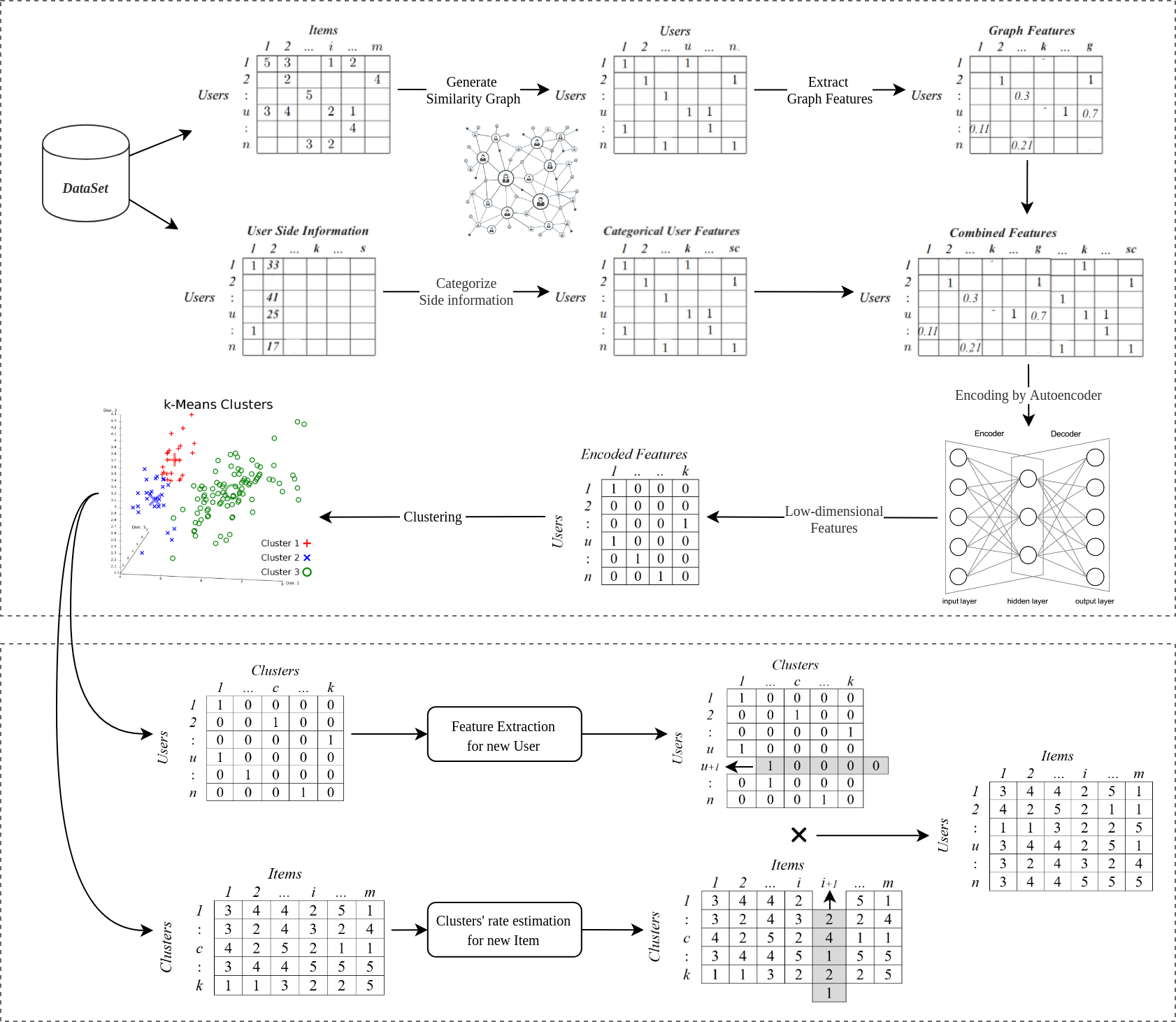}
	\caption{The framework of the proposed recommendation system. The method encodes the combined features with autoencoder and creates the model by clustering the users using the encoded features (upper part). At last, a preference-based ranking model is used to retrieve the predicted movie rank for the target user (lower part)}
	\label{FIG:03}
\end{figure*}

The overall structure of our aggregated recommender system (GHRS) is presented in Figure \ref{FIG:03}. The Graph-based Hybrid Recommender System comprises the following seven steps:
\begin{enumerate}
\item
In the first step, we build a graph with the number of users' as nodes. Two users will be connected based on their similarities. The edge connects a pair of users who have more than $\alpha$ percent of items with similar ratings.
\item
In the second step, a set of information will be extracted from the similarity graph for each user. For instance, we compute PageRank of the nodes, degree centrality, closeness centrality, the shortest-path betweenness centrality, load centrality, and the average degree of each node's neighborhood in the graph. As a result, this matrix relies on the different data processing magnitude using a preference-based collaborative approach.
\item
In the third step, we combine side information such as gender and age with graph-based features to retrieve the most relevant movies for users. Therefore, we have one combined matrix from different types of features, which is then used as the Autoencoder stage input.
\item
In the fourth step, we apply the Autoencoder to extract new features and reduce the dimension. It includes selecting a proper optimizer, using a proper loss function and neural network architecture, and preventing the overfitting issue.
\item
In the fifth step, we utilize the new features encoded by Autoencoder for user clustering, using the K-means algorithm to create a small number of peer groups. It includes finding an appropriate number of clusters for each dataset.
\item
In the sixth step, we assign new users to clusters based on encoded features and compute the new item rating base on similarity with other items.
\item
In the seventh step, we compute the estimated rates of all items for each user according to its cluster's average rating..
\end{enumerate}
  
Algorithm \ref{ALG:01} declared the total workflow in details.

\begin{algorithm}
	\caption{Proposed method detailed workflow}
	\label{ALG:01}

	\hspace*{\algorithmicindent} \textbf{Input:} $U, I, R, F_u, F_i$ \\
	\hspace*{\algorithmicindent} \textbf{Output:} Estimated rates for user-item

	\begin{algorithmic}[1]

		\State Set alpha = percentage of items with similar ratings between two users
		\State Compute the aggregated similarity between users base of $\alpha$ (the percentage of items which two users rated them similarly)
		\State Construct the Similarity Graph and consider the users as nodes
		\State $F_g$: Extracted graph-based features for users (nodes)
		\State $F_s$: Preprocessed and categorized users' side information (demographic informations)
		\State Combine $F_g$ and $F_s$ in a single feature vector $F_t$.
		Apply the Autoencoder on $F_t$ and train the model with the best settings
		\State Encode the $F_t$ using the Autoencoder an extract the low dimensional feature vector $F_e$
		\State Find the optimum clusters for clustering users with $F_e$
		\State Perform user clustering using extracted features vector $F_e$ and find clusters $C$
		\State Generate the user-cluster matrix $UC$
		\State Estimate clusters' ratings for items matrix $CI$:

		\If{there are users rated the item i before in the cluster $c$}
		\State $CI_{ci}$ = average (users' rates of the item $i$ in the cluster $c$)
		\ElsIf{there are Similar Items to the item $i$, rated by users in the cluster $c$}
		\State $CI_{ci}$ = average (users' rates of Similar Items in the cluster $c$)
		\Else
		\State $CI_{ci}$ = average (all users' rates in the cluster $c$)
		\EndIf

		\State Estimate users' ratings' matrix $R' = UC \times CI$
		\State Compute the recommendation list for target user $u$

	\end{algorithmic}

\end{algorithm}

\subsection{Graph-based Features}
\label{gbfeatures}
This section reviews the intuition of some graph features that represent similarities and their general computational process.
\begin{itemize}

	\item \textbf{Page Rank}: Page Rank is an algorithm that measures the transitive influence or connectivity of nodes. It was initially designed as an algorithm to rank web pages \citep{54xing2004a}. We can compute the Page Rank by either iteratively distributing one node's rank (based on the degree) over the neighbors or randomly traversing the graph and counting the frequency of hitting each node during these paths. In this paper, we used the first method.

	\item \textbf{Degree Centrality}: Degree centrality measures the number of incoming and outgoing relationships from a node. The Degree Centrality algorithm can be used to find the popularity of individual nodes \citep{46freeman1979a}. The degree centrality values are normalized by dividing by the maximum possible degree in a simple graph n-1, where n is the number of nodes.

	\item \textbf{Closeness Centrality}: Closeness centrality is a way to detect nodes that can spread information efficiently through a graph \citep{46freeman1979a}. The closeness centrality of a node u measures its average farness (inverse distance) to all n-1 other nodes. Since the sum of distances depends on the number of graph nodes, closeness is normalized by the sum of the minimum possible distances $n-1$.

	      \begin{equation}
		      C_C(u) = \frac{n-1}{\sum_{v=1}^{n-1}d(v,u))}
	      \end{equation}

	      where $d(v,u)$ is the shortest-path distance between $v$ and $u$, and $n$ is the number of nodes in the graph. Nodes with a high closeness score have the shortest distances to all other nodes.

	\item \textbf{Betweenness Centrality}: Betweenness centrality is a factor which we use to detect the amount of influence a node has over the flow of information in a graph. It is often used to find nodes that serve as a bridge from one part of a graph to another \citep{55moore1959a}. The Betweenness Centrality algorithm calculates the shortest (weighted) path between every pair of nodes in a connected graph, using the breadth-first search algorithm \citep{55moore1959a}. Each node receives a score, based on the number of these shortest paths that pass through the node. Nodes that most frequently lie on these shortest paths will have a higher betweenness centrality score.

	      \begin{equation}
		      C_B(u) = \frac{\sigma(s,t|u)}{\sum_{s,t\in V}^{}\sigma(s,t)}
	      \end{equation}

	      where $V$ is the set of nodes, $\sigma(s,t)$ is the number of shortest path between $(s, t)$, and $\sigma(s,t|u)$ is the number of those paths passing through some node $u$ other than $s$ and $t$. If $s = t$ $\rightarrow$ $\sigma(s,t)=1$, and if $v\in\{s,t\}$ $\rightarrow$ $\sigma(s,t|u)=0$ \citep{47brandes2008a}.

	\item Load Centrality. The load centrality of a node is the fraction of all shortest paths that pass through that node \citep{48newman2001a}.

	\item Average Neighbor Degree. Returns the average degree of the neighborhood of each node. The average degree of a node i is:

	      \begin{equation}
		      AND(u)=\frac{1}{N(u)}\sum_{v\in N(u)}^{}k_v
	      \end{equation}

	      where $N(u)$ are the neighbors of node $u$ and $k_v$ is the degree of node $v$ which belongs to $N(u)$. For weighted graphs, an analogous measure can be defined \citep{49barrat2004a},

	      \begin{equation}
		      AND^w(u)=\frac{1}{S_u}\sum_{v\in N(u)}^{}w_{uv}k_v
	      \end{equation}

	      where $s_u$ is the weighted degree of node $u$, $w_{uv}$ is the weight of the edge that links $u$ and $v$, and $N(u)$ are the neighbors of node $u$.

\end{itemize}

\subsection{User Clustering}
\label{clustering}
As we mentioned before in section \ref{architecture}, each user belongs to a specific cluster and the cluster rate for an item will be considered as the estimated rating for the user-item pair. In the proposed method we use K-Mean algorithm to cluster the users based on extracted features by Autoencoder. One important issue in using such algorithms is to find the proper number of clusters regarding performance factors. We use two methods to choose the number of clusters; Elbow method and Average Silhouette algorithm.

In this section we will explain the summary of K-Mean algorithms and how we tackle and solve the number of clusters issue with both mentioned methods.

\subsubsection{K-Means Algorithm}
\label{kmeans}
The K-means algorithm is a simple iterative clustering algorithm. Using the distance as the metric and given the K classes in the data set, calculate the distance mean, giving the initial centroid, with each class described by the centroid \citep{57yuan2019a, 60m2015a}. For a given data set X with n data samples and the number of category K, the Euclidean distance is the measure of the similarity index, and the clustering method aims minimize the sum of the squares of the various types. It means that it minimizes \citep{58wang2012a}
\begin{equation}
	d=\sum_{k=1}^{K}\sum_{i=1}^{n}\left \| (x_i-u_k) \right \|^2
\end{equation}

where $k$ represents $K$ cluster centers, $u_k$ represents the $k^{th}$ center, and $x_i$ represents the $i^{th}$ point in the data set.

\subsubsection{Elbow Method}
The basic idea behind cluster partitioning methods, such as k-means clustering, is to define clusters such that the total intra-cluster variation (known as a total within-cluster variation or total within-cluster sum of squares) is minimized. It measures the compactness of the clusters, and it should be as small as possible \citep{51kaufman2009a}. The elbow method is based on plotting the explained variation as a function of the number of clusters, and picking the elbow of the output curve as the proper number of clusters. Adding another cluster after the the elbow point doesn't give much better modeling of the data and may causes over-fitting.

\subsubsection{Average Silhouette}
Briefly, the average silhouette approach measures the quality of a clustering. It means that it determines how well each object occupies within its cluster. A high average silhouette width intimates a valuable clustering.

The average silhouette method computes the average silhouette of observations for different values of k. The optimal number of clusters k is the one that maximizes the average silhouette over a range of possible values for $k$ \citep{51kaufman2009a}.

\section{Empirical Experiments and Performance Evaluation}
\label{testandres}
In this section, the performance of the proposed model is evaluated, analyzed, and enumerated in separate parts. The dataset is processed and described in detail, followed by the requisite experimental setup. Due to the variation of steps and processes in the proposed method, we elaborate on the practical results in-depth. Finally, we compared the method with basics and modern methods, which we discussed most of them in related works.

\subsection{Dataset}
\label{dataset}
We have utilized two benchmark datasets (MovieLens 100K and MovieLens 1M) of the real-world in recommender systems to implement the model practically \citep{61harper2015a}. MovieLens 100K contains 100,000 ratings $R \in \{1, 2, 3, 4, 5\}$, 1,682 movies (items) rated by 943 users. MovieLens 1M comprises of 1,000,209 ratings $R \in \{1, 2, 3, 4, 5\}$ of approximately 3,900 movies made by 6,040 users. As discussed in Section \ref{ghrs}, the proposed method uses users' demographic data to solve the new users' cold-start issue. Hence, due to the lack of users' demographic data in larger datasets like MovieLens 10M, it would not be possible to evaluate the model more on larger datasets. We used the MovieLens 100K dataset for analyzing the proposed method's steps. The final evaluations and comparisons have been done on the MovieLense 1M dataset. Table \ref{TBL:01} shows the details of the mentioned datasets.

\begin{table*}[width=1\textwidth,cols=7,pos=h]
	\caption{Details of the datasets used for evaluation}\label{TBL:01}
	\begin{tabular*}{\tblwidth}{@{} LLLLLLL@{} }
		\toprule
		Dataset & Users & Items & Ratings & Rating Scale & Density & Source\\
		\midrule
		MovieLens 100K & 943 & 1,682 & 100,000 & [1-5] & 6.304\% & \citep{61harper2015a} \\
		MovieLens 1M & 6,040 & 3,900 & 1,000,209 & [1-5] & 1.431\% & \citep{61harper2015a}\\
		\bottomrule
	\end{tabular*}
\end{table*}

\subsection{Features Statistics}
\label{statistics}
As declared in Section \ref{ghrs}, we use two types of features in the proposed method: side information (users' demographic data) and features extracted from the similarity graph between users. We transformed the demographic data into a categorical format, concatenated both types of features, and made the raw feature set before dimension reduction with an Autoencoder. In this section, we discuss a little about the statistics of the raw features.

\begin{figure*}

	\centering
	\begin{subfigure}{0.23\textwidth}
		\centering
		\includegraphics[width=\linewidth]{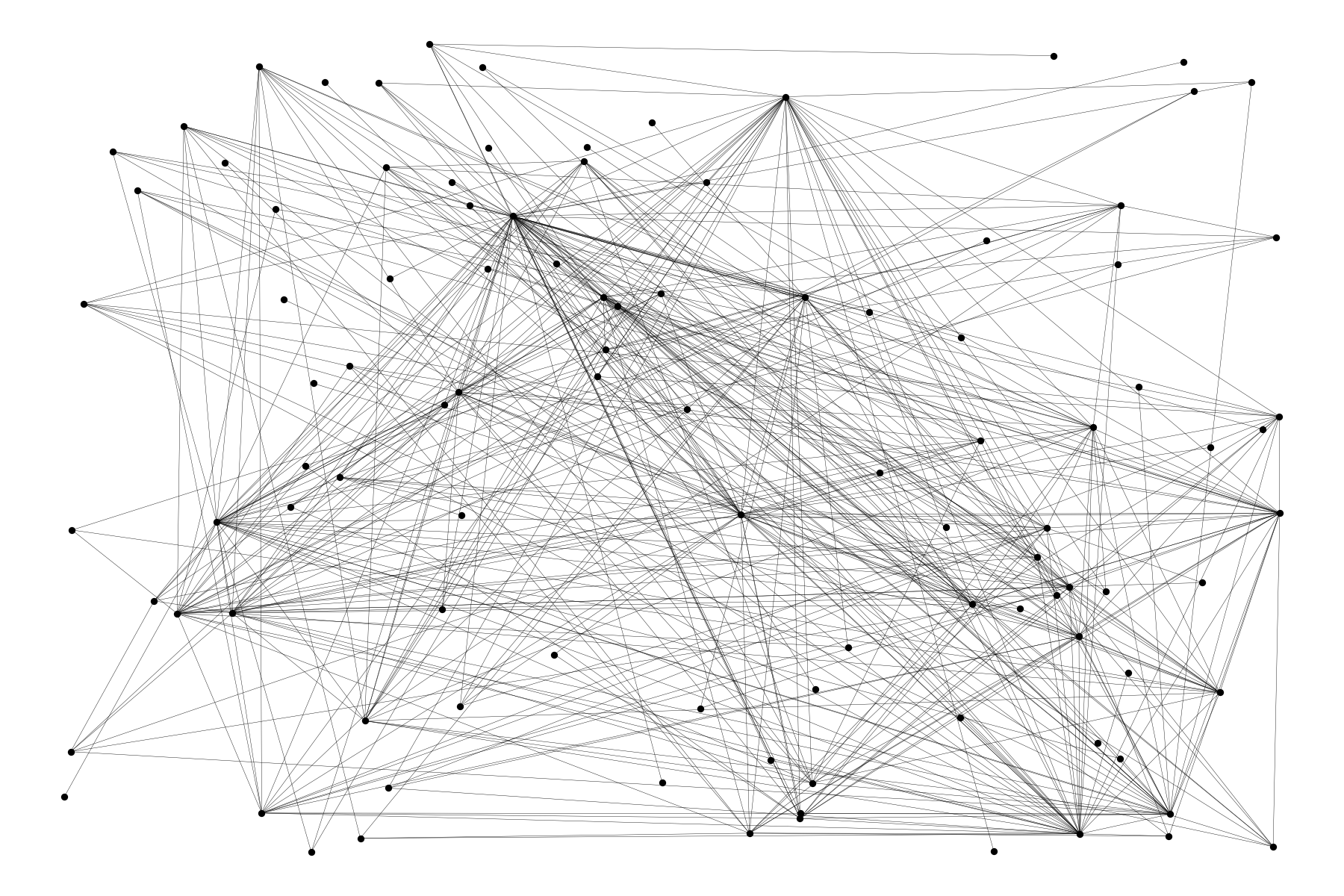}
		\caption{$\alpha=0.03$}
		\label{FIG:04:a}
	\end{subfigure}
	\begin{subfigure}{0.23\textwidth}
		\centering
		\includegraphics[width=\linewidth]{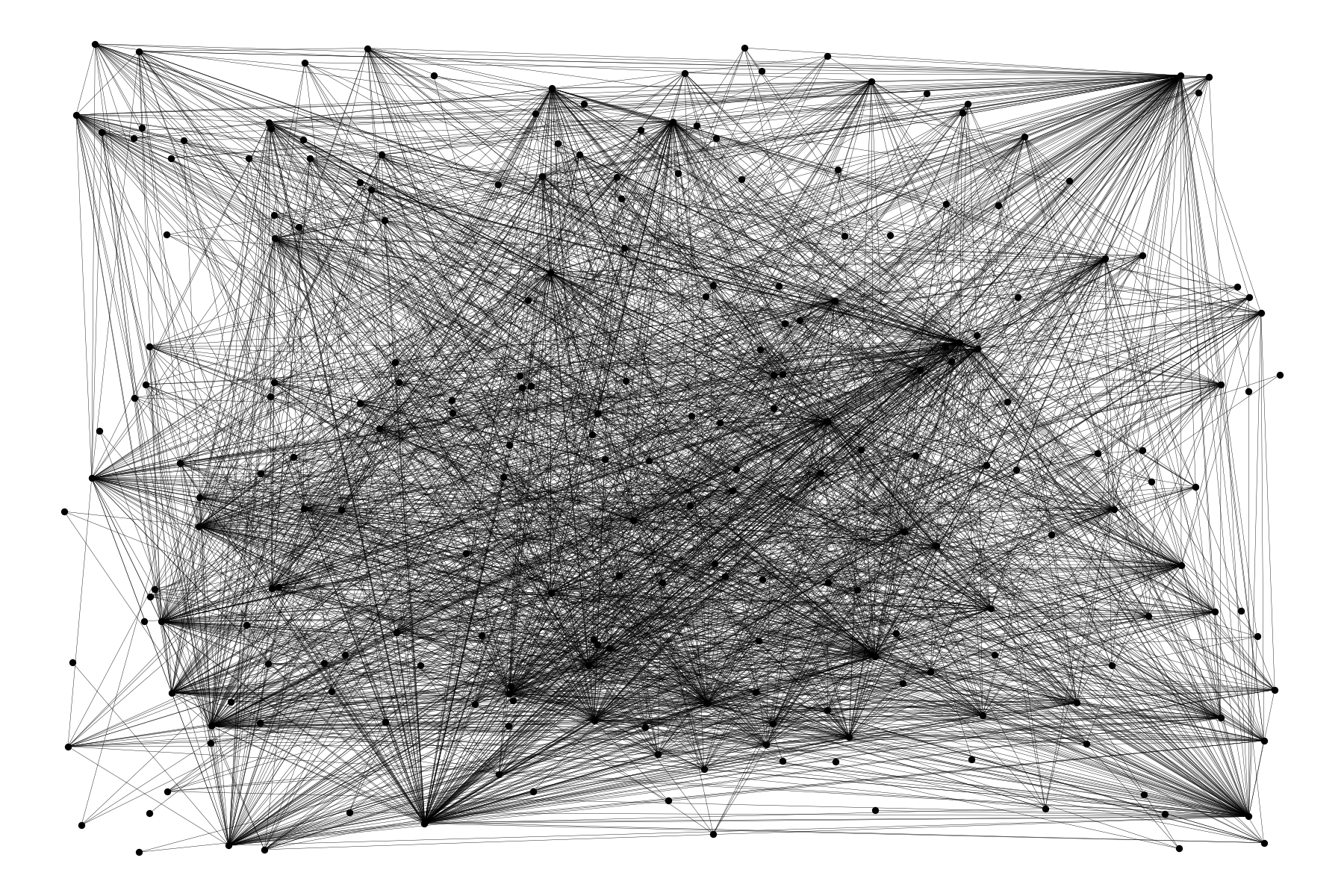}
		\caption{$\alpha=0.02$}
		\label{FIG:04:b}
	\end{subfigure}
	\begin{subfigure}{0.23\textwidth}
		\centering
		\includegraphics[width=\linewidth]{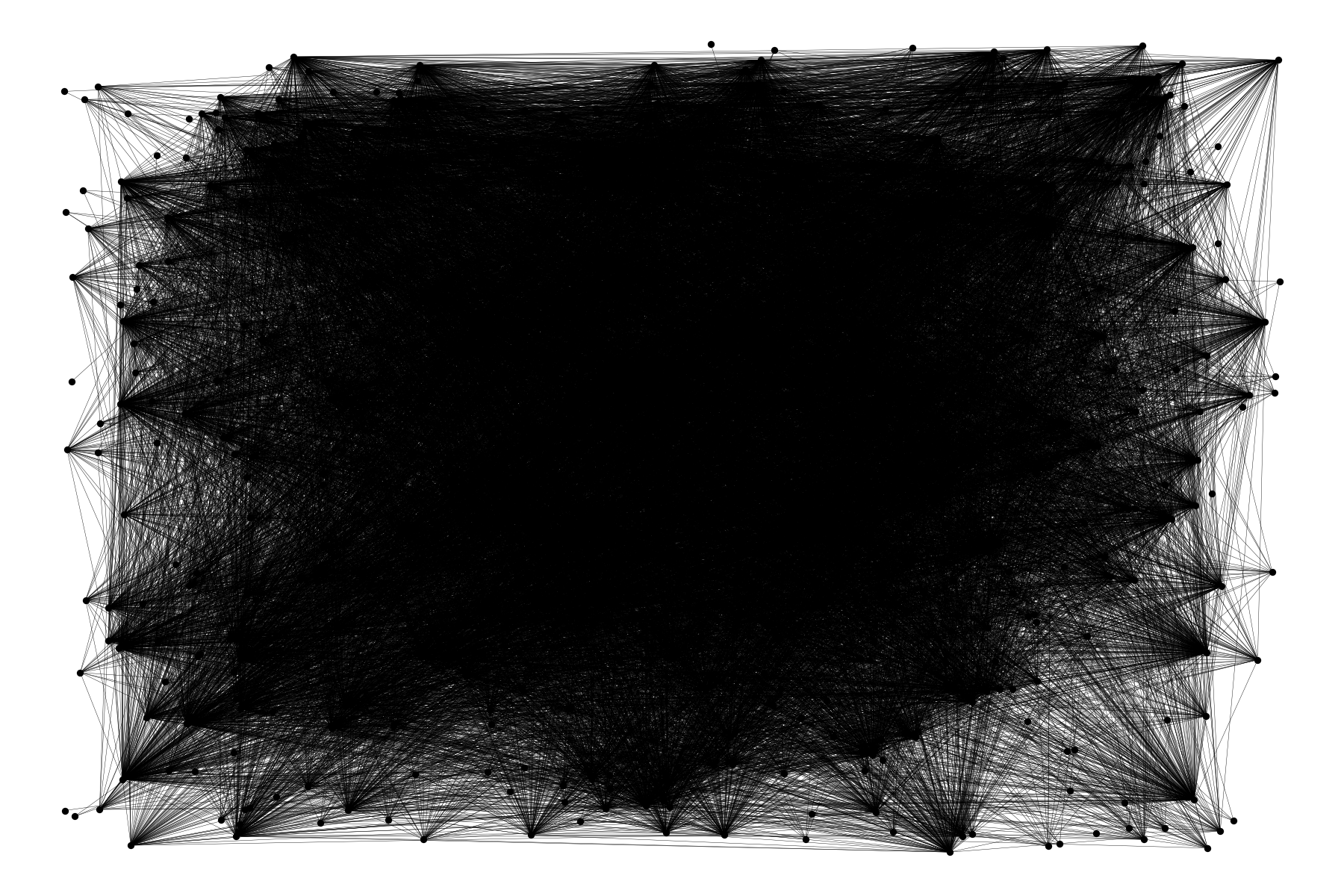}
		\caption{$\alpha=0.01$}
		\label{FIG:04:c}
	\end{subfigure}
	\begin{subfigure}{0.23\textwidth}
		\centering
		\includegraphics[width=\linewidth]{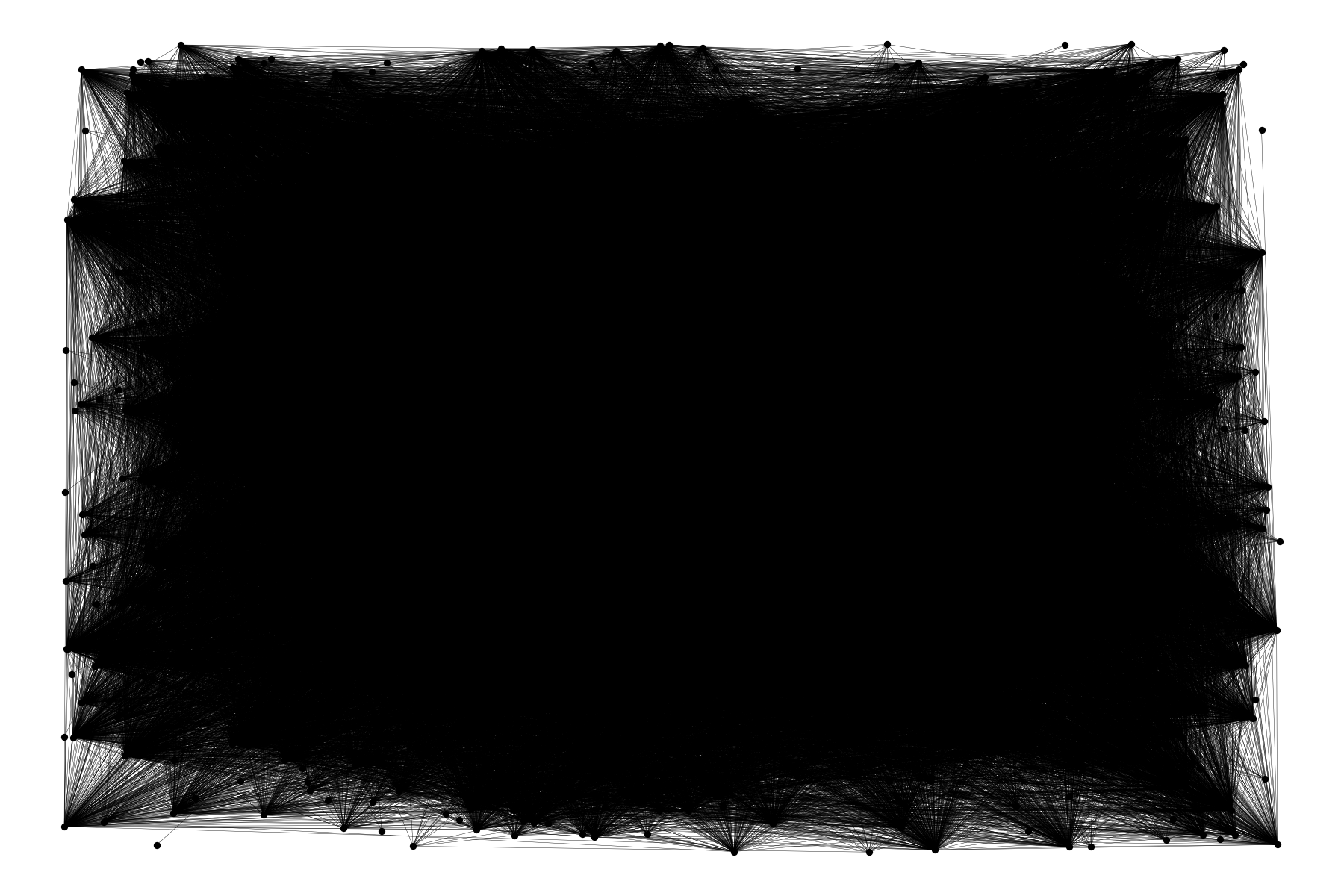}
		\caption{$\alpha=0.005$}
		\label{FIG:04:d}
	\end{subfigure}

	\caption{Visualization of Similarity Graph for $\alpha=\{0.005, 0.01, 0.02, 0.03\}$ for MovieLens 100K with 943 users.}
	\label{FIG:04}
\end{figure*}

We have declared before the only parameter we have used for generating the graph is $\alpha$, the value of a threshold for connecting two users having at least several same movies in their ratings. This threshold is represented as a percentage of total movies in the dataset. Hence, we have an exploration of a very sparse graph to near a full-mesh graph. Figure \ref{FIG:04} illustrates the similarity graph visualization for $\alpha=\{0.005, 0.01, 0.02, 0.03\}$ for 943 users in MovieLens 100K.

\begin{figure}

	\begin{subfigure}{.45\textwidth}
		\centering
		\includegraphics[width=0.9\linewidth]{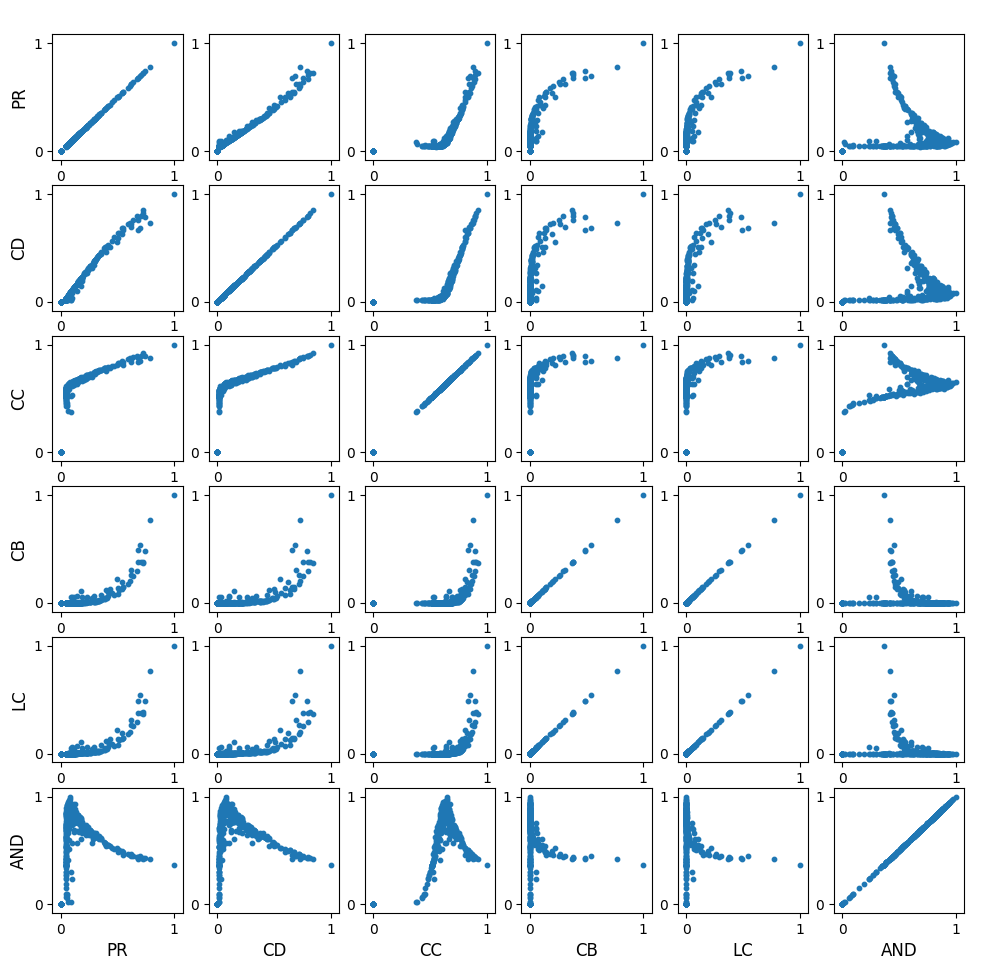}
		\caption{MovieLens 100K}
		\label{FIG:05:a}
	\end{subfigure}
	\begin{subfigure}{.45\textwidth}
		\centering
		\includegraphics[width=0.9\linewidth]{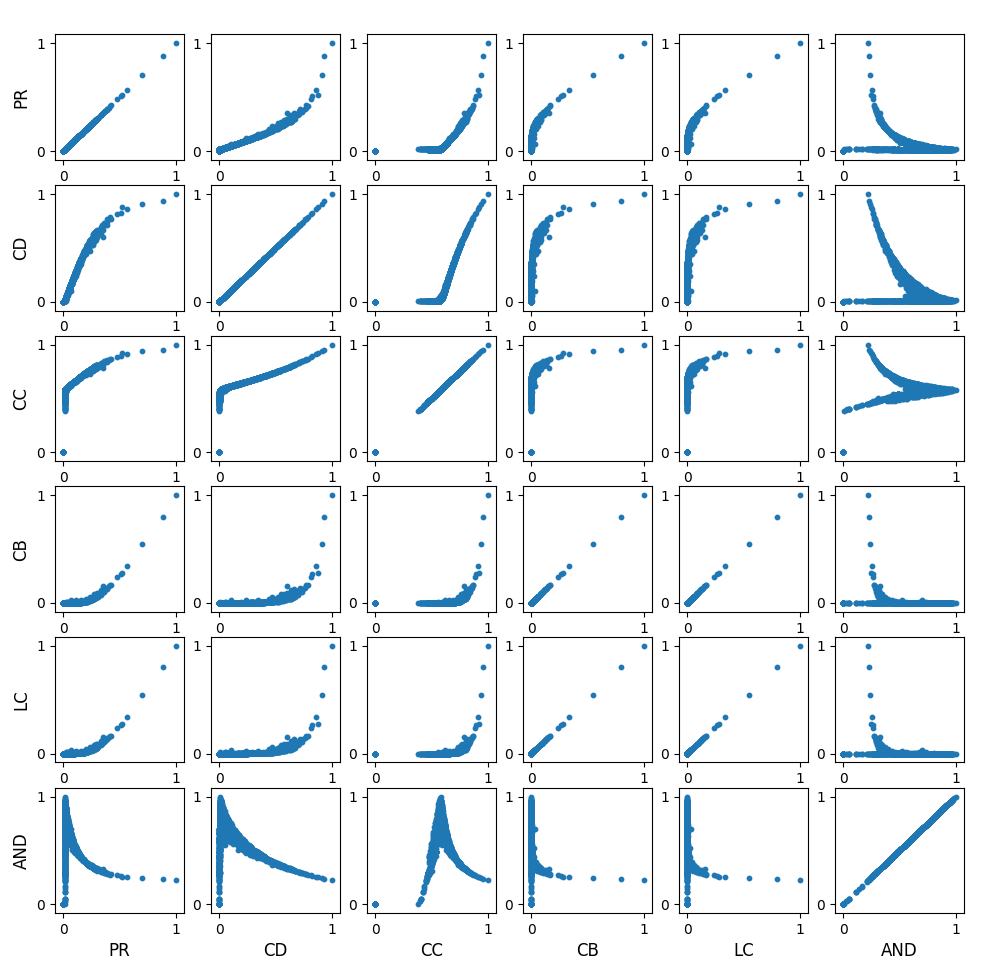}
		\caption{MovieLens 1M}
		\label{FIG:05:b}
	\end{subfigure}

	\caption{Graph-based features for (a) MovieLens 100K and (b) MovieLens 1M both for $\alpha=0.015$. Abbreviation used in the figure: PR (page rank), CD (degree centrality), CC (closeness centrality), CB (betweenness centrality), AND (average neighbor degree), and LC (load centrality). Centrality measures like CD and PR have more correlation than others, and it may be caused by the correlation of eigenvector centrality and degree centrality in our user graph. It means that users with a high degree of connection are likely connected in many cases.}
	\label{FIG:05}
\end{figure}

Figure \ref{FIG:05} shows the normalized graph-based features' distributions against each other for MovieLens 100K and MovieLens 1M with $\alpha=0.015$. We can see correlations between these types of features in some cases.

\begin{figure}
	\centering
	\includegraphics[scale=.5]{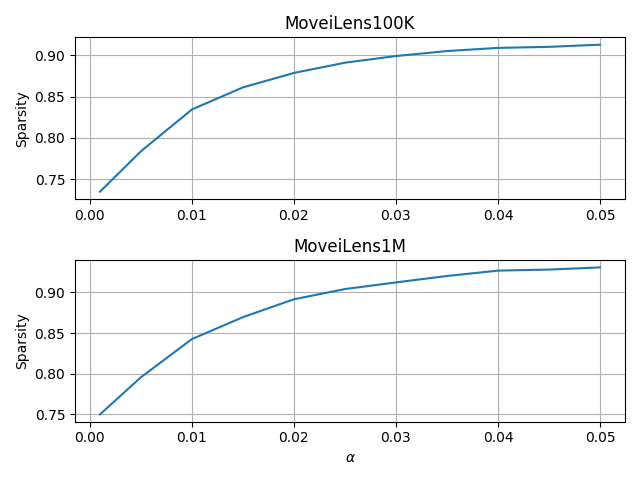}
	\caption{Sparsity of combined features' dataset \textit{vs.} $\alpha$ before dimension reduction}
	\label{FIG:06}
\end{figure}

As all the demographic features are transformed into a categorical format, the demographic features vector is one-hot encoded and has a specific sparsity level for each dataset. On the other hand, we declared that the graph-based features' value is related to the similarity graph size, and the graph size is directly related to the factor $\alpha$. In Figure \ref{FIG:06}, we can see that the feature set's sparsity rises when the value of the $\alpha$ increases.

\subsection{Performance Metrics}
\label{metrics}
We use 10-fold cross-validation on MovieLens 1M dataset and 5-fold cross-validation on MovieLens 100K dataset to partition the datasets into training and testing to measure the performance of the GHRS. The final prediction metrics are the average of the iterations of training and testing base on the number of folds in each dataset. The training set comprises the User-Item list with given ratings, user's demographic information, and item's side information. We consider the Root Mean Squared Error (RMSE) as the metric for evaluation. RMSE (Equation \ref{eq_RMSE}) is generally related to the cost function of conventional rating prediction models:

\begin{equation}
	RMSE = \sqrt{\frac{\sum_{u,i}^{u \in U, i \in I} (R_{ui} - R'_{ui})}{Number of Ratings}}
	\label{eq_RMSE}
\end{equation}

Where $U$ is user set, $I$ is item set, and $R_{ui}$ and $R'_{ui}$ are the actual and predicted ratings of user $u$ for item $i$, respectively.

Besides those mentioned above, we also use Precision and Recall (the most popular metrics for evaluating information retrieval systems) as an evaluation metric to measure the proposed model's accuracy. Precision measures the ratio of correct recommendations to the total recommendations, and Recall shows the ratio of correct recommendations to total correct information. Consequently, we have to separate the items into two classes with a threshold while considering their actual ratings, i.e., non-relevant and relevant to measure Precision and Recall. In this regard, items rated between $[1-3]$ were considered non-relevant and rated with $[4-5]$ as relevant. Additionally, the items in datasets were divided as selected and not selected on their predicted ratings. Therefore, Precision and Recall of the model can be defined as:
\begin{equation}
	Precision = \frac{TP}{TP+TF}
\end{equation}
\begin{equation}
	Recall = \frac{TP}{TP+FN}
\end{equation}
Where $TP$ stands for True Positive (Item is correctly selected as it is relevant), $FP$ stands for False Positive (Item is incorrectly selected as it is not relevant), and $FN$ stands for False Negative (Relevant item is not selected)\\

Root Mean Square Error (RMSE) for the proposed model gives a lower error value for the testing dataset. Observed results produced by iterations of cross-validation (10-fold for MovieLens 1M and 5-fold for MovieLens 100K) show almost similar validation errors.

\subsection{Impact of similarity graph size}
\label{test:simgraph}
\begin{figure}
	\centering
	\includegraphics[scale=0.5]{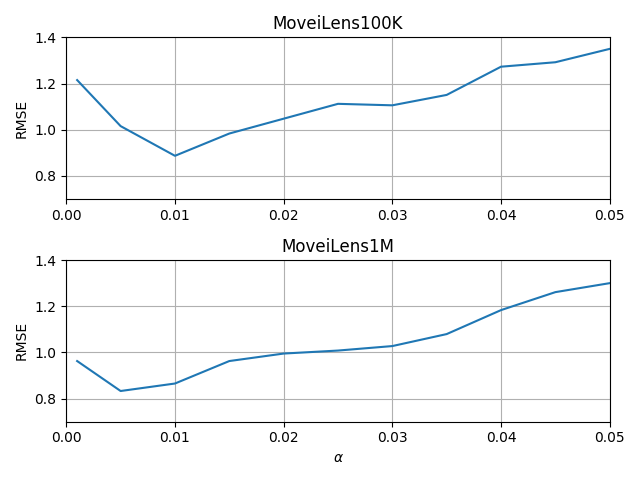}
	\caption{RMSE \textit{vs.} $\alpha$. There is an optimum point for $\alpha$ near the alpha = 0.01 for dataset MovieLens 100K and near the alpha = 0.005 for MovieLens 1M.}
	\label{FIG:07}
\end{figure}

In this experiment, we check the impact of graph size on rating accuracy. As we use graph features for every node (users) in the similarity graph, it's important to produce a similarity graph in a state that represents the similarity between nodes as optimized as it can be. For this purpose, we experimented with searching in parameter space, which impacts the size and the shape of the similarity graph. Figure \ref{FIG:07} shows the RMSE vs. $\alpha$ in both dataset we used for the evaluation. As it can be seen in the figure, there is no direct relation between the result of the method. But, the minimum value of RMSE achieved on a specific value of alpha in the middle of the experiment range.

The main reason for this result is that when the alpha's value is very small, all users can be connected due to this value because we consider just a very little common items in their ratings to connect them to each other in the similarity graph. Hence, most of the users are similar to each other in this condition and the difference will be missed in some cases. On the other hand, when the alpha's value raises the similarity graph become more sparse (As it is shown in Figure \ref{FIG:06}). So, we consider the most of users not related to each other when the $\alpha$ value increases to very large values. There is an optimum point for the size of the similarity graph near the alpha = 0.01 for dataset MovieLens 100K and near the alpha = 0.005 for MovieLens 1M. The result of this parameter tuning has been produced with the exact condition of the final evaluation with k-fold cross-validation (Figure \ref{FIG:06}).

\subsection{Dimension Reduction}
\label{test:dr}
\begin{figure}
	\centering
	\includegraphics[width=\linewidth]{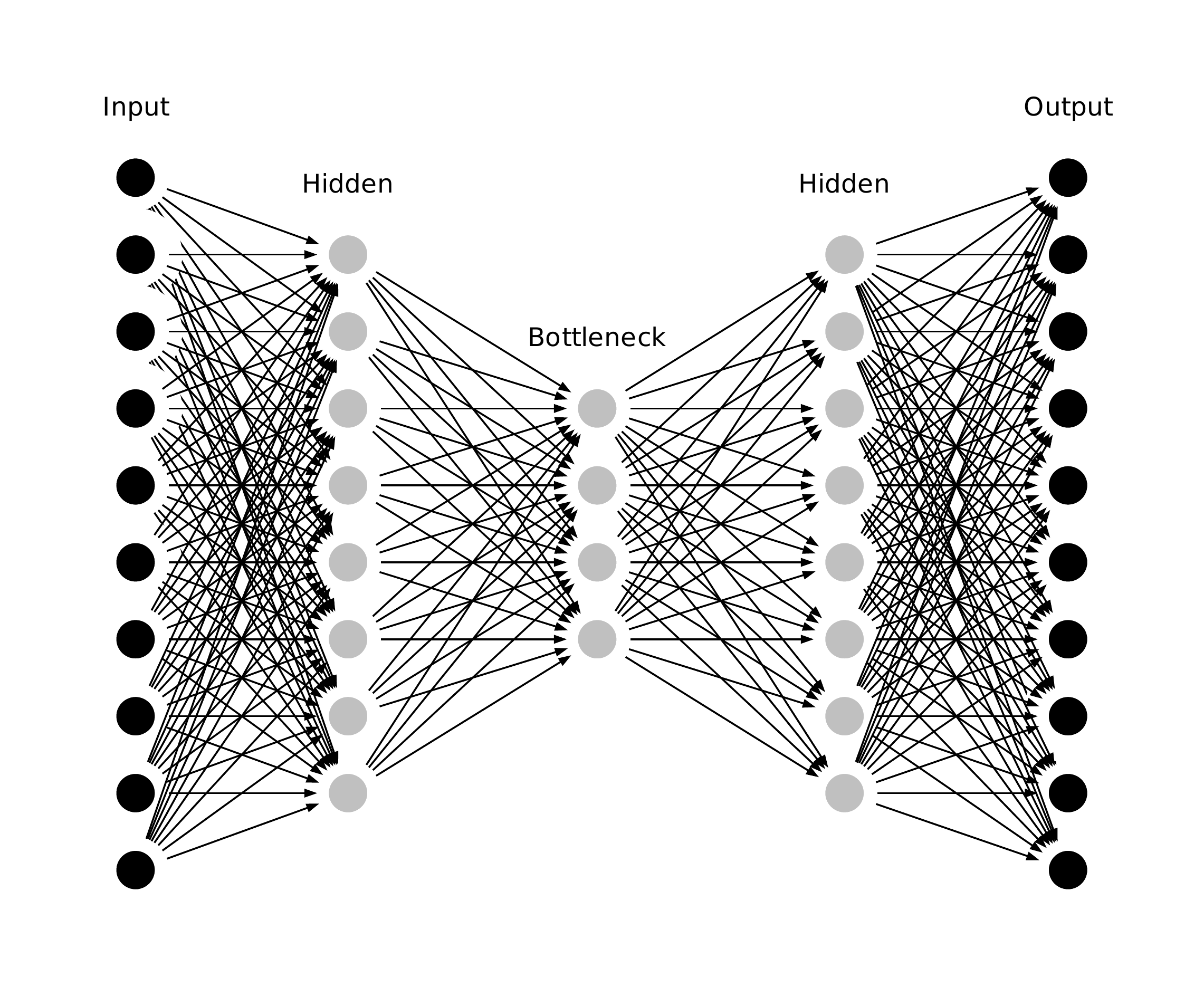}
	\caption{Structure of Autoencoder used for dimension reduction. Both input and output size are 35 for MovieLens 100K and 36 for MovieLens 1M}
	\label{FIG:08}
\end{figure}

We declared that we use Autoencoder to simultaneously extract new features and reduce the raw feature set dimension before clustering. In this experiment, we examine the learning algorithms for Autoencoder and check each algorithm's ability to minimize the loss function on our input raw feature set. In all experiments, we have used a 5-layer Autoencoder with the structure shown in Figure \ref{FIG:08}. The activation function of all layers is ReLU \citep{62nair2010rectified}.

Both input and output size (raw feature vector's sizes) are 35 for MovieLens 100K and 36 for MovieLens 1M.

We experiment with a set of optimizers to train the Autoencoder. Optimizers include Adagrad \citep{63duchi2011adaptive}, Adadelta \citep{64zeiler2012adadelta}, RMSProp \citep{65hinton2012neural}, Adam \citep{66kingma2014adam}, AdaMax \citep{66kingma2014adam}, Nadam \citep{67dozat2016incorporating} and SGD \citep{86bottou2008tradeoffs} with loss function of Mean Squared Error.

\begin{figure}
	\centering
	\includegraphics[scale=0.5]{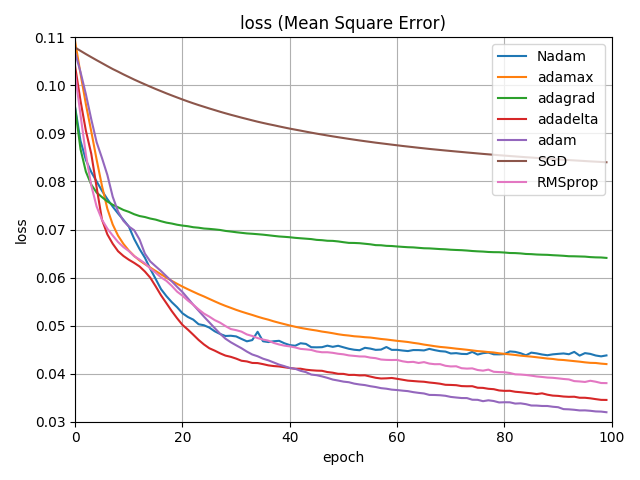}
	\caption{The value of loss function on validation data over 100 epochs of training with seven target optimizers in the experiment}
	\label{FIG:09}
\end{figure}

Figure \ref{FIG:09} shows the result of the experiment and compares the optimizers in minimizing the loss function on validation data over 100 epochs of training. We have randomly selected 10\% of users in MovieLens 1M and 20\% of users in MovieLens 100K as validation data and exclude them from the training set. We can see the best results for Adam, Adadelta, and RMSProp optimizers. As a discussion about the result, RMSprop can be considered as an extension of Adagrad that deals with its radically diminishing learning rates. It is identical to Adadelta, except that Adadelta uses the RMS of parameter updates in the nominator update rule. Adam, finally, adds bias-correction and momentum to RMSprop \citep{69ruder2016overview}.

RMSprop, Adadelta, and Adam are very similar algorithms that do well in similar conditions. \citet{66kingma2014adam} showed that the bias-correction helps Adam slightly exceed RMSprop towards the end of optimization when the gradients become sparser. Hence, Adam seems to be the best option for the optimizer.  Recently, many researchers use vanilla SGD without momentum and a simple learning rate annealing schedule \citep{69ruder2016overview}. Nevertheless, In our experiment, SGD approaches to achieves a minimum, but it may take longer than other methods.

We selected Adam as the optimizer in Autoencoder to encode the raw feature set. The output of the encoding process shows a diverse distribution with a low correlation between the encoded features. Figure \ref{FIG:11} shows the encoded features for both MovieLens 100K and MovieLens 1M, which will be used for clustering the users.

We use elastic net regularization (linear combination of $L_1$ and $L_2$ penalties) \citep{87zou2005regularization} for Autoencoder to avoid the overfitting on the training data and improve the model's performance.

\begin{figure}
	\centering

	\begin{subfigure}{0.25\textwidth}
		\centering
		\includegraphics[width=0.9\linewidth]{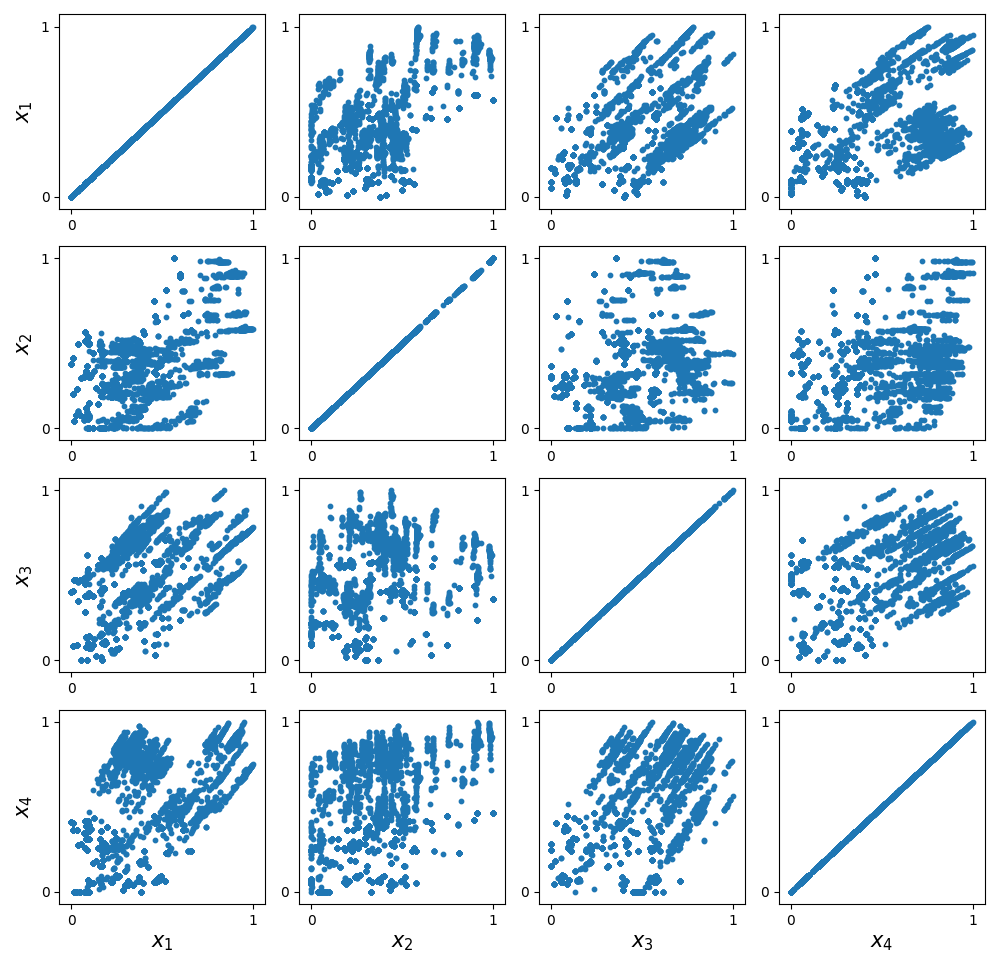}
		\caption{MovieLens 1M}
		\label{fig10:sfig1}
	\end{subfigure}%
	\begin{subfigure}{0.25\textwidth}
		\centering
		\includegraphics[width=0.9\linewidth]{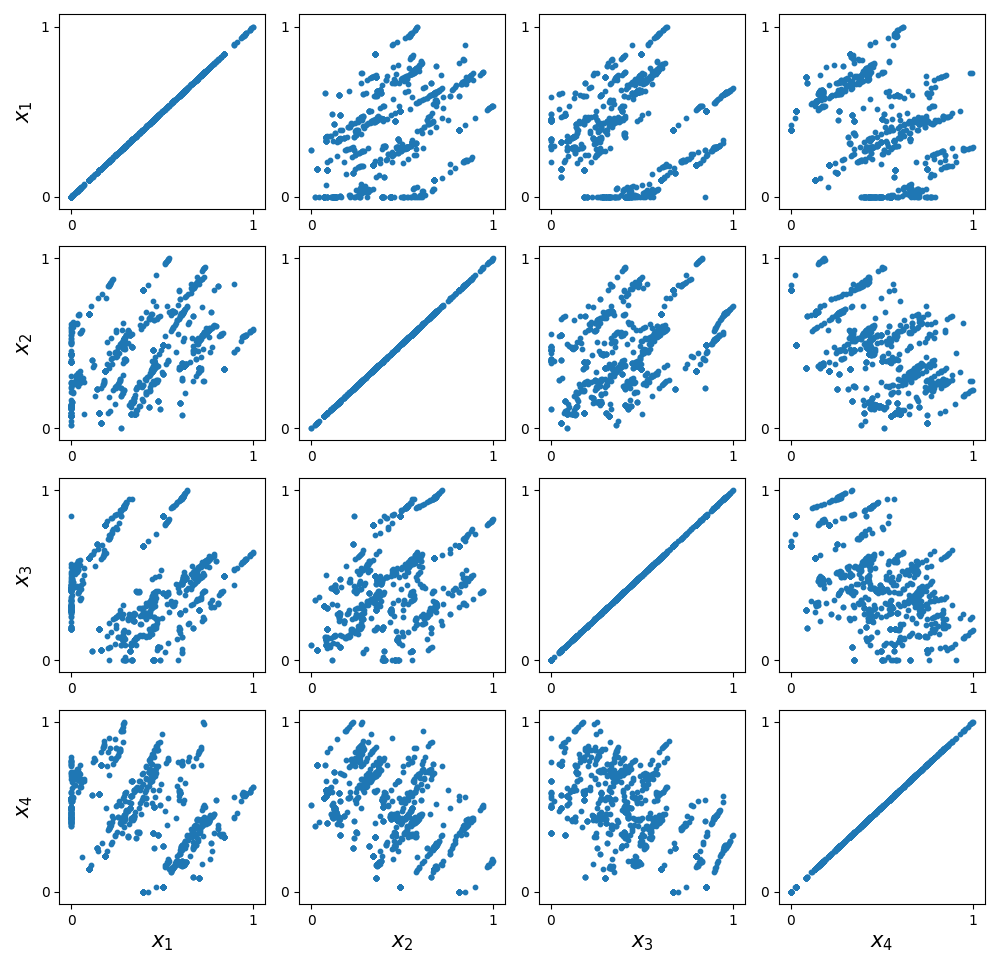}
		\caption{MovieLens 100K}
		\label{fig10:sfig2}
	\end{subfigure}

	\caption{Final Features Scatter (Dimension is reduced to 4 using Autoencoder with Adam optimizer).}
	\label{FIG:10}
\end{figure}

\subsection{Number of Clusters}
\label{test:clustering}
\begin{figure}

	\begin{subfigure}{.5\textwidth}
		\centering
		\includegraphics[width=1\linewidth]{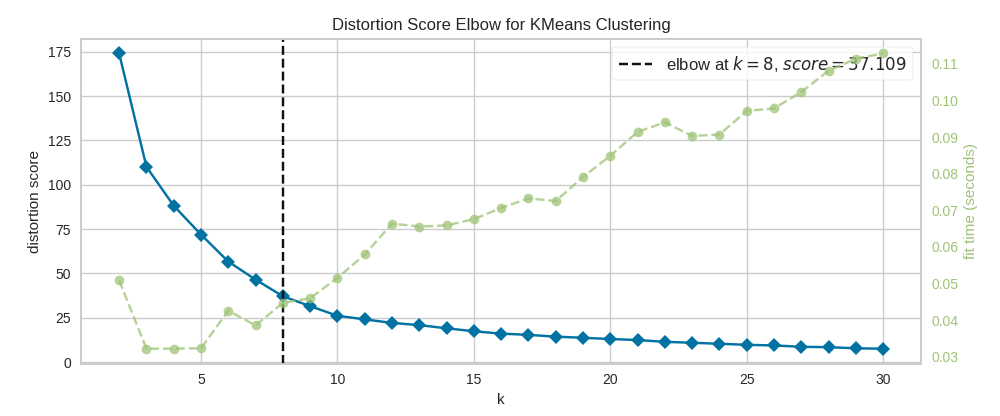}
		\caption{MovieLens 100K - Elbow Method on K-Mean}
		\label{fig:sfig1}
	\end{subfigure}
	\begin{subfigure}{.5\textwidth}
		\centering
		\includegraphics[width=1\linewidth]{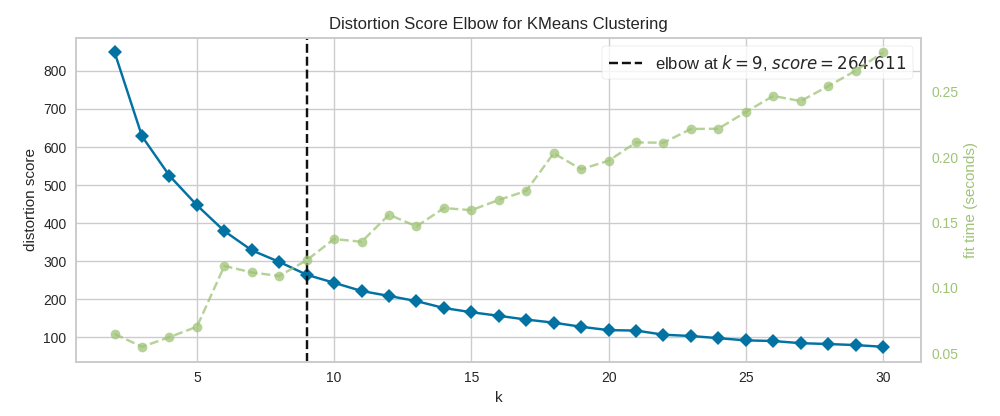}
		\caption{MovieLens 1M - Elbow Method on K-Mean}
		\label{fig:sfig1}
	\end{subfigure}

	\caption{Finding the elbow point as the optimum number of clusters of users. (a) K-Mean Algorithm. (b) MiniBatchK-Mean Algorithm}
	\label{FIG:11}
\end{figure}

\begin{figure}

	\begin{subfigure}{.5\textwidth}
		\centering
		\includegraphics[width=1\linewidth]{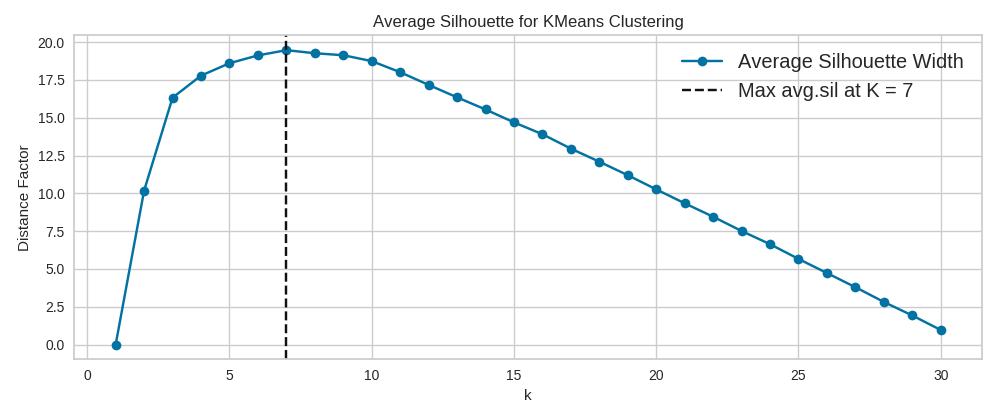}
		\caption{MovieLens 100K - Average Silhouette on K-Mean}
		\label{fig:sfig1}
	\end{subfigure}
	\begin{subfigure}{.5\textwidth}
		\centering
		\includegraphics[width=1\linewidth]{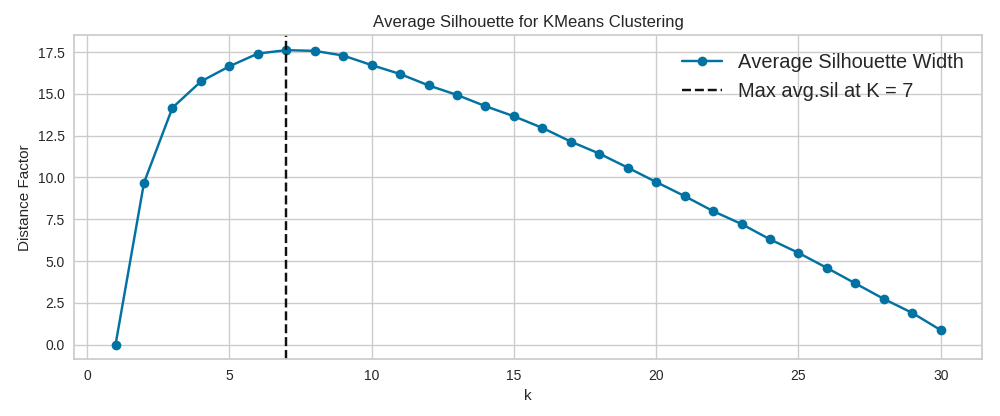}
		\caption{MovieLens 1M - Average Silhouette on K-Mean}
		\label{fig:sfig1}
	\end{subfigure}

	\caption{Finding the elbow point as the optimum number of clusters of users. (a) K-Mean Algorithm. (b) MiniBatchK-Mean Algorithm}
	\label{FIG:12}
\end{figure}

This section examines the mentioned method in section \ref{clustering} to find the correct number of clusters for both datasets MovieLens 1M and MovieLens 100K. As listed before, we applied two methods for this reason; the Elbow method and the Average Silhouette method. The input of both methods is the encoded feature sets from the previous state. For both methods, we consider the range of K in $[1-30]$. Figure \ref{FIG:11} and Figure \ref{FIG:12} show the algorithms' iteration for the Elbow method and Average Silhouette method, respectively. The best value of K has been founded, as shown in Table \ref{TBL:02}.

\begin{table*}[width=.8\textwidth,cols=3,pos=h]
	\centering
	\caption{Value of K suggested by Elbow method and Average Silhouette method}\label{TBL:02}
	\begin{tabular*}{\tblwidth}{@{} LLL@{} }
		\toprule
		Dataset & Elbow method & Average Silhouette method\\
		\midrule
		MovieLens 100K & $K=8$, $Distortion \ Score = 37.11$ & $K=7$, $Average \ Silhouette = 18.93$ \\
		MovieLens 1M & $K=9$, $Distortion \ Score = 264.61$ & $K=7$, $Average \ Silhouette = 17.53$\\
		\bottomrule
	\end{tabular*}
\end{table*}

\subsection{Results and Comparison with Other Methods}
\label{test:res}
Performance of the proposed model has been evaluated on the datasets mentioned in section \ref{dataset}. Table \ref{TBL:03} shows the result of the proposed model based on the best setting derived from the experiments conducted to find the best values for parameters $\alpha$ (section \ref{test:simgraph}) and K (section \ref{test:clustering}), and best optimizer for Autoencoder (section \ref{test:dr}).

\begin{table}[width=\linewidth,cols=4,pos=h]
	\centering
	\caption{Performance metrics value for the proposed method on target dataset}\label{TBL:03}
	\begin{tabular*}{\tblwidth}{llll}
		\toprule
		Dataset & RMSE & Precision & Recall \\
		\midrule
		100K & 0.887, $S^2$=$1.595\times 10^{-4}$ & 0.771 & 0.799 \\
		1M & 0.833, $S^2$=$2.815\times 10^{-4}$ & 0.792 & 0.838 \\
		\bottomrule
	\end{tabular*}
\end{table}

\begin{figure}
	\centering
	\includegraphics[scale=0.52]{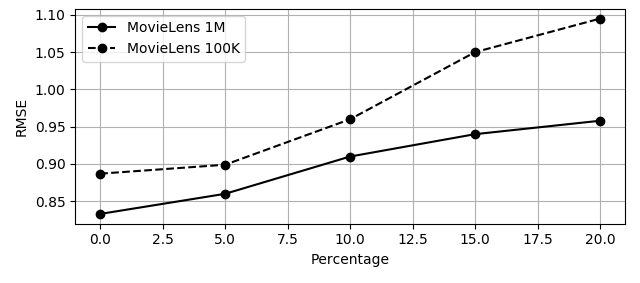}
	\caption{GHRS method RMSE result versus the percentage of users which have been randomly removed from the user-item rating matrix.}
	\label{FIG:13}
\end{figure}

In another experiment, we are going to assess the GHRS method in tackling the cold-start problem. For this reason, we had to produce a synthetic dataset from the original dataset like MovieLens 100k or 1M. In the synthetic dataset, regardless of the number of ratings, we have randomly removed a specific percentage of records from matrix $R$ (user-item ratings) before generating graph-based features for each user. Hence, it has the same condition as a new user without any previous rating records, and we have only side information to predict ratings for this user. Figure \ref{FIG:13} shows the method performance result versus the percentage of users which have been randomly removed from the user-item rating matrix. We can see that GHRS delivers a good performance in cold-start issues, even in a high percentage of new users.

The proposed method (GHRS) has additionally been compared with some primary methods and state of the art methods. It is evident that the proposed method shows an improvement in the best result of RMSE on MovieLens 1M and has the best performance as same as AutoRec after the Autoencoder COFILS. Full comparison is shown in Table \ref{TBL:04}.

\begin{table*}[width=\textwidth,cols=3,pos=h]
	\centering
	\caption{Comparison with other methods}\label{TBL:04}
	\begin{tabular*}{\tblwidth}{lll}
		\toprule
		Model & MovieLens 100K & MovieLens 1M \\
		\midrule
		Collaborative Topic Regression \citep{70wang2011a} & - & 0.896 \\
		Collaborative Deep Learning \citep{34wang2015b} & - & 0.887 \\
		Convolutional Matrix Factorization \citep{72kim2016a} & - & 0.853 \\
		Convolutional Matrix Factorization+ \citep{72kim2016a} & - & 0.854 \\
		Robust Convolutional Matrix Factorization \citep{73kim2017a} & - & 0.847 \\
		RippleNet \citep{74wang2018a} & - & 0.863 \\
		Imputed Singular Value Decomposition \citep{75yuan2018a} & - & 0.85 \\
		Genetic Algorithm and Gravitational Emulation \citep{76mohammadpoura2019a} & -	 & 1.087 \\
		Noise Correction Based RS \citep{77bag2019a} & - & 1.7 \\
		DST-HRS \citep{78khan2020a} & - & 0.846 \\
		Autoencoder COFILS \citep{85barbieri2017autoencoders} & \textbf{0.885} & 0.838 \\
		Baseline COFILS \citep{85barbieri2017autoencoders} & 0.892 & 0.848 \\
		Kernel PCA COFILS \citep{85barbieri2017autoencoders} & 0.898 & - \\
		Slope One \citep{79lemire2005slope} & 0.937 & 0.9 \\
		Regularized SVD \citep{80paterek2007improving} & 0.989 & 0.96 \\
		Improved Regularized SVD \citep{80paterek2007improving} & 0.954 & 0.907 \\
		SVD++ \citep{81koren2008factorization} & 0.903 & 0.856 \\
		Non-Negative Matrix Factorization \citep{82lee2001algorithms} & 0.944 & 0.912 \\
		Bayesian Probabilistic Matrix Factorization \citep{15salakhutdinov2008a} & 0.901 & 0.84 \\
		RBM-CF \citep{83salakhutdinov2007restricted} & 0.936 & 0.872 \\
		AutoRec \citep{25sedhain2015a} & 0.887 & 0.844 \\
		Mean Field \citep{84langseth2015scalable} & 0.903 & 0.856 \\
		GHRS (Proposed Method) & 0.887 & \textbf{0.833} \\
		\bottomrule
	\end{tabular*}
\end{table*}

\section{Conclusion and Future Works}
\label{conclusion}
We have proposed a method for the recommendation in user-item systems in this paper. The method can be used for every user-item system that provides side information for both users and items. The proposed method's main idea is finding the relation between users based on their similarities as nodes in a similarity graph and combining them with the users' side information to solve the cold-start issue. Plus, we applied Autoencoder to extract new low dimensional features with low correlation and more information. This made the final clustering step more accurate and highly performed in time consumption. Final experiments and comparison with other methods showed the competitive results for the selected datasets and improved the best result on MovieLens 1M dataset.

There are several lines of research arising from this work that should be pursued. Future research might apply for the work on the item properties like user side information to detect similarity between items precisely. Admittedly, it will be like considering similarities between two users who similarly rate the same items, and their rates and properties in the similarity graph are close to each other. Indeed, in this case, items will be considered similar if identical or similar users (based on similarity definition between users in this research) rate them with the same patterns. Thus, we will also have the approach using graph features and deep-learning for users, for items.

On the other hand, it is great to devote future research to developing and extracting more other features from the similarity graph, which we did not mention in the current study.
Besides, the structure of the Autoencoder might be an important area for future research. The different structures should be examined regarding the Autoencoder structure affecting feature extraction, training duration, and the model's final performance. In this article, we used a predefined structure for Autoencoder using the heuristic method and manual tuning. Also, there are many methods to cluster the users in this method. They should be investigated and measured to find the optimal one for this type of feature space and distribution. This assumptions might be addressed in future studies.

As discussed in section \ref{dataset}, few datasets have side information for users and items(e.g., demographic data for users). It will be desirable to assess the proposed method with other future datasets that include this information.

\printcredits
\bibliography{cas-refs}

\bibliographystyle{cas-model2-names}

%
%

\end{document}